\documentclass[twocolumn,tighten]{aastex631}
\usepackage{xspace}
\usepackage{eucal}
\usepackage{float}
\usepackage{graphicx} 
\usepackage[multiple]{footmisc}
\usepackage{amsmath, esint}
\usepackage{amssymb}

\def\lsim{\mathrel{\raise.3ex\hbox{$<$\kern-.75em\lower1ex\hbox{$\sim$}}}}
\def\gsim{\mathrel{\raise.3ex\hbox{$>$\kern-.75em\lower1ex\hbox{$\sim$}}}}

\defcitealias{EHTC_M87_I}{\m87~I}
\defcitealias{EHTC_M87_II}{\m87~II}
\defcitealias{EHTC_M87_III}{\m87~III}
\defcitealias{EHTC_M87_IV}{\m87~IV}
\defcitealias{EHTC_M87_V}{\m87~V}
\defcitealias{EHTC_M87_VI}{\m87~VI}
\defcitealias{EHTC_M87_VII}{\m87~VII}
\defcitealias{EHTC_M87_VIII}{\m87~VIII}
\defcitealias{EHTC_M87_IX}{\m87~IX}
\defcitealias{EHTC_M87_2018_I}{\m87-2018-I}
\defcitealias{EHTC_SgrA_I}{\sgra~I}
\defcitealias{EHTC_SgrA_II}{\sgra~II}
\defcitealias{EHTC_SgrA_III}{\sgra~III}
\defcitealias{EHTC_SgrA_IV}{\sgra~IV}
\defcitealias{EHTC_SgrA_V}{\sgra~V}

\defcitealias{EHTC_SgrA_VI}{\sgra~VI}
\defcitealias{EHTC_SgrA_VII}{\sgra~VII}
\defcitealias{EHTC_SgrA_VIII}{\sgra~VIII}

\def\m87{M87$^*$\xspace}
\def\sgra{Sgr~A$^*$\xspace}

\begin{document}
\title{
Assessing the Role of Intrinsic Variability in Black Hole Parameter Inference using Multi-Epoch EHT Data
}
\correspondingauthor{Dominic O. Chang}
\email{dochang@g.harvard.edu}

\author[0000-0001-9939-5257]{Dominic O. Chang}
\affiliation{Department of Physics, Harvard University, Cambridge, Massachusetts 02138, USA}
\affiliation{Black Hole Initiative at Harvard University, 20 Garden Street, Cambridge, MA 02138, USA}

\author[0000-0002-4120-3029]{Michael D. Johnson}
\affiliation{Black Hole Initiative at Harvard University, 20 Garden Street, Cambridge, MA 02138, USA}
\affiliation{Center for Astrophysics $|$ Harvard \& Smithsonian, 60 Garden Street, Cambridge, MA 02138, USA}

\author[0000-0003-3826-5648]{Paul Tiede}
\affiliation{Center for Astrophysics $|$ Harvard \& Smithsonian, 60 Garden Street, Cambridge, MA 02138, USA}

\begin{abstract}
Event Horizon Telescope (EHT) observations of M87* provide a means of constraining parameters of both the black hole and its surrounding plasma. 
However, intrinsic variability of the emitting material introduces major sources of uncertainty which complicates parameter inference. 
The precise nature of this variability remains uncertain, and previous studies have largely relied on general relativistic magnetohydrodynamic (GRMHD) simulations to estimate its effects. Here, we fit a semi-analytic, dual-cone model of the emitting plasma to multiple years of EHT observations to empirically assess the impact of intrinsic variability and improved array coverage on key measurements including the black hole mass-to-distance ratio, spin, and viewing inclination. 
Despite substantial differences in the images of the two epochs, we find that the inferred mass-to-distance ratio remains stable and mutually consistent. 
The black hole spin is unconstrained for both observations, despite the improved baseline coverage in 2018. 
We show that intrinsic variability can contribute significantly to the inference error, and that the inferred position angle and inclination of the black hole spin axis are discrepant between the two years. 
Our findings highlight both the promise and challenges of multi-epoch EHT observations: while they can refine parameter constraints, they also reveal the limitations of simple parametric models in capturing the full source complexity. 
Our analysis -- the first to fit semi-analytic emission models to 2018 EHT observations -- underscores the importance of  quantifying data contributions from intrinsic variability in future high-resolution imaging studies of black hole environments and the role of repeated observations in quantifying these uncertainties.
\end{abstract}

\section{Introduction}

Black holes are the central engines of Active Galactic Nuclei (AGN), giving rise to observable structures seen over many orders of magnitude in scale \citep{agnjetreview}. Many of these phenomena are engendered by processes that originate on sub-parsec scales and propagate out to influence larger structures. 
Structures on the smallest scales likely occur near the event horizon of the central supermassive black hole. Here, strong relativistic effects lead to image features that encode information about the black hole and its environment, and these can now be studied directly using images with very long baseline interferometry (VLBI). 
Strong-field gravitational lensing, for example, gives rise to a characteristic dark depression -- the black hole's apparent shadow -- whose size can be used to determine the mass-to-distance ratio ($\theta_{\rm g}$) of the black hole \citep{Falcke_2000,EHTC_M87_VI,EHTC_SgrA_IV}. This inference provides an independent and complementary approach to stellar and gas dynamical modeling techniques \citep[e.g.,][]{Gebhardt, Walsh, Jeter_2021,Liepold_2023, Simon}.

\m87 has long been an ideal candidate for the study of AGN since its angular size and luminosity allow it to be studied over a vast range of length scales and across the entire electromagnetic spectrum \citep[e.g.,][]{EHT_M87_MWL_2021,EHT_M87_MWL_2024}. 
Recently, the near-horizon structure of \m87 has been imaged using observations with the Event Horizon Telescope (EHT) in 2017 and 2018 \citep[][]{m87I,m87II,m87III, m87IV,m87V,m87VI,M87_2018_I,M87_2018_II}.
With these observations, the EHT Collaboration (EHTC) estimated $\theta_{\rm g} = 3.62^{+0.41}_{-0.34}\;\mu$as from a combined data study over the two observed epochs \citep{M87_2018_II}.
\citet{Chang_2024} recently re-analyzed the 2017 EHT observations of \m87 using Bayesian inference techniques with a dual-cone synchrotron jet model, leading to estimates of $\theta_{\rm g}$ within the range of $(2.84,3.75)\;\mu$as to $95\%$ confidence. 
Both measurements are consistent with the stellar dynamics measurements of \m87 from \citet{Gebhardt}, $\theta_{\rm g} = 3.62\pm0.22\;\mu$as, and from \citet{Liepold_2023}, $\theta_{\rm g} = 3.16^{+0.22}_{-0.15}\;\mu$as.\footnote{Both works report mass measurements for \m87 that we have converted to $\theta_{\rm g}$ assuming a distance of $D = 16.8\,{\rm Mpc}$ \citep{m87VI}.}

The EHTC uses a geometric modeling technique to estimate $\theta_{\rm g}$, relating it to specific image features. In particular, the diameter of the observed ring is correlated with the mass of the black hole. However, it can also be sensitive to the observing resolution, black hole spin, and the geometry of the emitting region.
The EHTC quantified systematic uncertainties from these additional effects using a library of General Relativistic Magnetohydrodynamic (GRMHD) simulations and an associated ``$\alpha$-calibration'' procedure  \citep{EHTC_M87_VI,EHTC_SgrA_IV}. This procedure relates the ring diameter $d$ in each simulated image to the mass-to-distance ratio via a calibration factor, $\alpha = d/\theta_{\rm g}$. The distribution of $\alpha$ across a suite of simulations defines an approximation of the systematic uncertainty in this mapping, but a limitation of this procedure is its reliance on simulations. 
\citet{Chang_2024}, in contrast, employed a more direct procedure in which a simple dual-cone emission model is fit directly to the EHT data. This model includes parameters of the black hole such as $\theta_{\rm g}$ and those that describe the average emission properties. 
A major limitation of this procedure is that it does not include any estimates of the uncertainties arising from intrinsic source variability, which corresponds to misspecification for the underlying emission model.

Here, we extend the analysis of \citet{Chang_2024} to explore the effects of source variability on parameter inference by repeating the analysis for two independent EHT observations of \m87, in 2017 and 2018. Because the images in these two epochs are notably different, especially in a shift of the peak ring brightness, this approach provides an empirical estimate for the systematic uncertainty and bias arising from intrinsic variability. In \autoref{sec:model_desc}, we summarize the dual-cone model and define its parameters. 
In \autoref{sec:data_method}, we describe the EHT observations of \m87 in 2017 and 2018 used in this study and our approach to parameter inference using the EHT data. 
In \autoref{sec:results}, we summarize our results.

\section{Description of the model}
\label{sec:model_desc}

\begin{table*}[htbp]
\centering
\begin{center}
\setlength\extrarowheight{3pt}
\setlength{\tabcolsep}{3pt}
\begin{tabular}{c|c|l} 
 \hline \hline
 \multicolumn{3}{c}{Dual-Cone Model Parameters}\\
 \hline
 \textbf{Classification} & \textbf{Parameter} & \textbf{Description} \\ \hline
 & $\theta_{\rm g}$ & Mass-to-distance ratio ($\mu$as) \\
\textbf{Black Hole} & $a$ & Black hole dimensionless spin \\
\textbf{and Observer} & $\theta_{\rm o}$ & Observer inclination with respect to black hole spin axis  (deg.) \\
 & p.a. & Position angle of projected spin axis on observer's screen (deg.) \\
 \hline
& $\theta_s$ & Cone opening angle (deg.) \\
 & $R$ & Characteristic radius of intensity profile $(\text{GM}/c^2)$\\
 & $p_1$ & Inner exponent of intensity profile \\
\textbf{Accretion} & $p_2$ & Outer exponent of intensity profile \\
\textbf{and} & $\chi$ & Fluid velocity azimuthal angle in ZAMO frame (deg.)\\
\textbf{Emission} & $\iota$ & Magnetic field orthogonality angle in fluid frame (deg.)\\
& $\beta_v$ & Fluid speed in ZAMO frame ($c$) \\
 & $\sigma$ & Spectral index of emission \\
 & $\eta$ & Magnetic field tangential angle in fluid frame (deg.) \\
\hline
\end{tabular}
\end{center}
\caption{
Parameters of the dual-cone model. Four parameters define the black hole and the observer's position with respect to it. The remaining nine parameters describe the dynamics, emissivity, and magnetic field orientation in the emitting plasma near the black hole. For additional details, see \citet{Chang_2024}.
}
\label{tab:model_params}
\end{table*}

The observed emission from \m87 at millimeter wavelengths has been shown to be well described by simulations that combine GRMHD with general relativistic ray tracing and radiative transfer \citep{m87VI, m87VIII}. 
Under these assumptions, the emission from \m87 is well described as synchrotron emission from plasma within a few Schwarzschild radii of the black hole in the accretion disk and jet-launching region \citep[see, e.g.,][]{Dexter,m87VI, m87VIII}.

Although GRMHD simulations successfully reproduce a wide variety of phenomena seen in the radio and infrared, their computational expense limits their use in formal statistical analyses of EHT data. As an alternative, \citet{Chang_2024} introduced a semi-analytic, dual-cone model as a proxy for numerical GRMHD, which they show to be a good descriptor of the time-averaged accretion flow. This model is semi-analytic and differentiable, building upon a series of even simpler toy models that also showed success in reproducing key elements of GRMHD images \citep{Narayan_2021,Gelles,Palumbo_2022} and integrating analytic expressions for null geodesics in the Kerr spacetime \citep{GrallaLupsasca,Krang}. 
The dual-cone model approximates the emission geometry as a compact profile that is constrained to two symmetric cones in Boyer-Lindquist coordinates whose axes align with the black hole's spin axis. 
The apices of the two cones are located at the Boyer-Lindquist coordinate origin.\footnote{The symmetry of the geometry is such that the dual-cone model would have had the same embedding even if defined with respect to Kerr-Schild coordinates instead of Boyer-Lindquist.}
The dual-cone model is defined by 13 parameters. Of these, 4 parameters characterize the black hole and its orientation with respect to the distant observer, and the remaining 9 parameters describe the emission geometry and magnetic field orientation.
\autoref{tab:model_params} summarizes these parameters.

For \m87, observations in a single night effectively capture a single temporal snapshot of the accretion flow, which is expected to be highly variable. 
The source variability likely arises from turbulent structures that dynamically form in the black hole accretion flow. 
In GRMHD simulations, these structures typically result in a correlation timescale of $\gtrapprox 50\, \text{GM/c}^3$, \citep[see, for example ][]{Boris,Conroy}, which for \m87 is $\sim 20$ days. Since the dual-cone model is a representation of the average accretion flow, it is expected to have some degree of model misspecification when fitting these snapshots from intrinsic variability. 
Using mock EHT observations of GRMHD simulations of \m87, \citet{Chang_2024} found that measurements of the dual-cone model are variable due to the influence of different instantiations of the accretion flow, although the systematic uncertainty from intrinsic variability was typically smaller than the statistical uncertainty. 
In this paper, we instead focus on empirically testing this conclusion with multiple observations of \m87 that sample statistically independent realizations of the accretion flow.

\section{Data and Methodology}
\label{sec:data_method}

\begin{figure}
    \centering
    \includegraphics[width=\columnwidth]{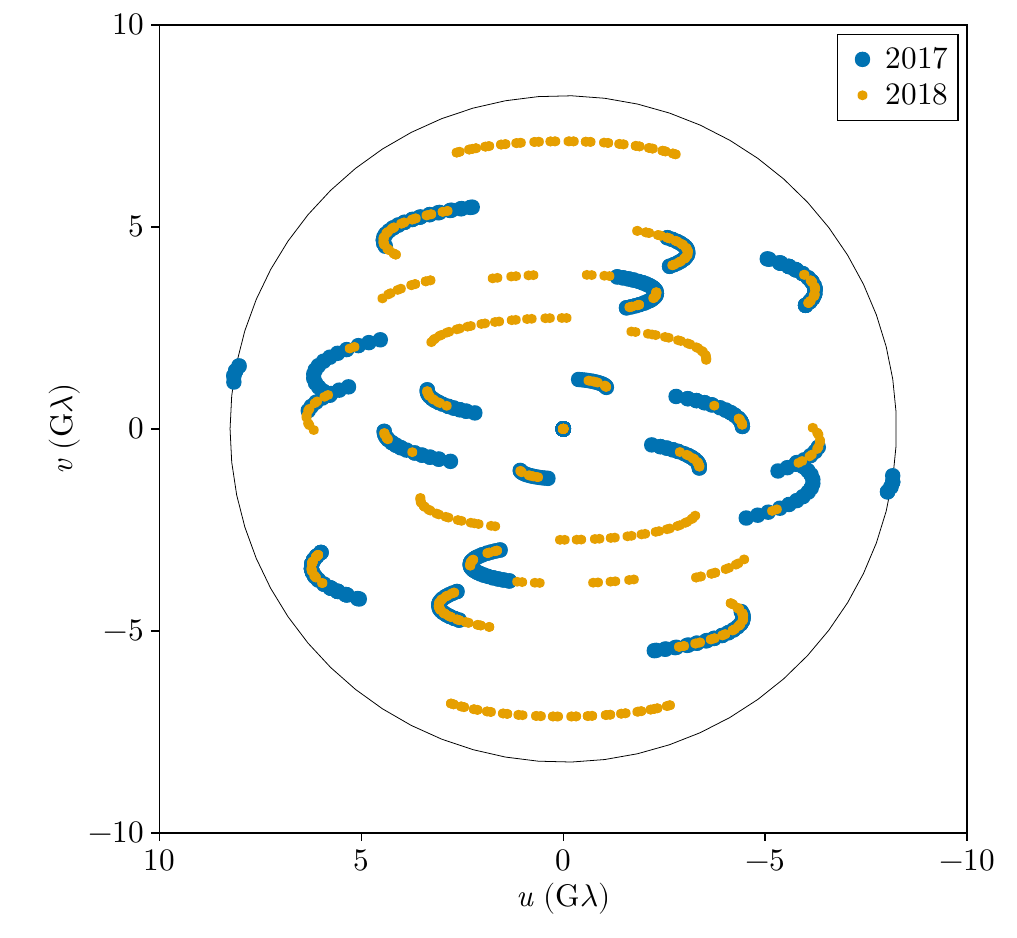}
    \caption{
    Baseline coverage $(u, v)$ from the EHT observations of \m87. EHT observations from April 6, 2017 and April 21, 2018 are shown in blue and orange, respectively. The gray circle shows a fringe spacing of $25\,\mu$as, corresponding to $\sqrt{u^2 + v^2} \approx 8.25\,{\rm G}\lambda$.
    }
    \label{fig:data-coverage}
\end{figure}

We fit the dual-cone model to EHT observations of \m87 in 2017 and 2018. For 2017, we used the data and fitting procedure described in \citet{Chang_2024}. For 2018, we used data from the 2\,GHz frequency band centered around 227.1 GHz \citep[called ``band 3" in ][]{M87_2018_I} that was acquired on April 21, 2018. This is the same band that was used in the \citet{Chang_2024} analysis of data from 2017 EHT observations of \m87 (called the ``low band" in the EHTC's analyses).

The 2018 observations included all sites in the 2017 campaign with the addition of the Greenland Telescope (GLT), resulting in a total of nine participating facilities. Thus, the data set in 2018 has more complete coverage than the data set in 2017 (see \autoref{fig:data-coverage}).

We construct our data products using the same procedure as \citet{Chang_2024}. 
In short, we 1) average data over ``scans,'' 2) exclude intra-site baselines (which are sensitive to flux in the large-scale jet, which is not part of the dual-cone model), and 3) add $1\%$ fractional noise to the resulting data products. 
The data have significant uncorrected calibration errors (commonly called ``gains''); rather than fitting these gains, we fit using log-closure amplitudes and closure phases, which are invariant to gain errors \citep{TMS}. To avoid biases from non-Gaussian closure statistics at low signal-to-noise (S/N), we only fit closure quantities that have ${\rm S/N} > 3$ \citep[see][for a description of closure quantity construction]{Blackburn}.

We perform a Bayesian inference study on the resulting data products with the dual-cone model, where we adopt the uniform priors as described in section~4 of \cite{Chang_2024}.
The prior on spin originates from theoretical analyses of the source by \citet{m87V}, which suggest the spin axis of \m87 points away from the Earth. 
We implement this prior by requiring the sign of the spin of the dual-cone model to be negative.
Our spin sign convention is consistent with one that associates the sign of the spin with its orientation relative to a distant observer \citep[see][]{Chang_2024, Palumbo_2022, Gelles}. 
In contrast, the convention adopted by the EHT corresponds to the relative orientation of the spin axis with respect to the fluid angular momentum.\footnote{The orientation of the spin axis is inferred from the southern location of the brightness peak in the EHT image of \m87. However, we find it possible for the dual-cone model to produce a southern brightness peak with a spin axis pointed towards the earth if the accreting fluids rotate with retrograde motion at sufficiently extreme relativistic speeds. Cursory analysis indicates that speeds $\gg0.99 c$ are required, as seen in the ZAMO frame.}
We take our prior on the cone opening to span the range, $\theta_s\in (40^\circ, 90^\circ)$.
This range covers the values seen in the image domain fits of the dual-cone model to a set of GRMHD simulations \citep[for details, see][]{Chang_2024}.
Additional comments on the chosen priors can be found in \citep{Chang_2024}.

We use a Gaussian likelihood for the visibility data products:
\begin{align}
    \mathcal L(\mathbf{ q}|\hat{\mathbf q})
        &
        \propto 
        \exp\left[
            -\frac{1}{2}\tilde{\mathbf q}^\dagger\mathbf\Sigma_q^{-1}\tilde{\mathbf q}
        \right].
        \label{eqn:likelihood}
\end{align}
Here, $\mathbf \Sigma_q$ is the covariance matrix, $\mathbf q$ are the measured visibility data products, $\hat{\mathbf q}$ are the data products from the model assumption, and,
\begin{align}
    \tilde{\mathbf q}
        &=\mathbf q-\hat{\mathbf q},
\end{align}
are the data product residuals. When the data products are closure phases $\mathbf \Psi$, or log-closure amplitudes $\mathbf c$ then,
\begin{align}
    \mathbf q
        &=\begin{cases} 
        \exp(i\mathbf\Psi)
        &\text{closure phases}\\
        \mathbf c
        &\text{log-closure amplitudes}.
        \end{cases}
\end{align}

We use the VLBI statistical inference framework \texttt{Comrade.jl} \citep{Tiede2022}, and sample the posterior with the \texttt{Julia} \citep{julia} implementation of a non-reversible parallel tempering algorithm \citep{surjanovic2023pigeonsjl} with the default slice-sampler \citep{Neal}.

\section{Results}
\label{sec:results}

\begin{figure}[htbp]
    \centering
    \includegraphics[width=\linewidth]{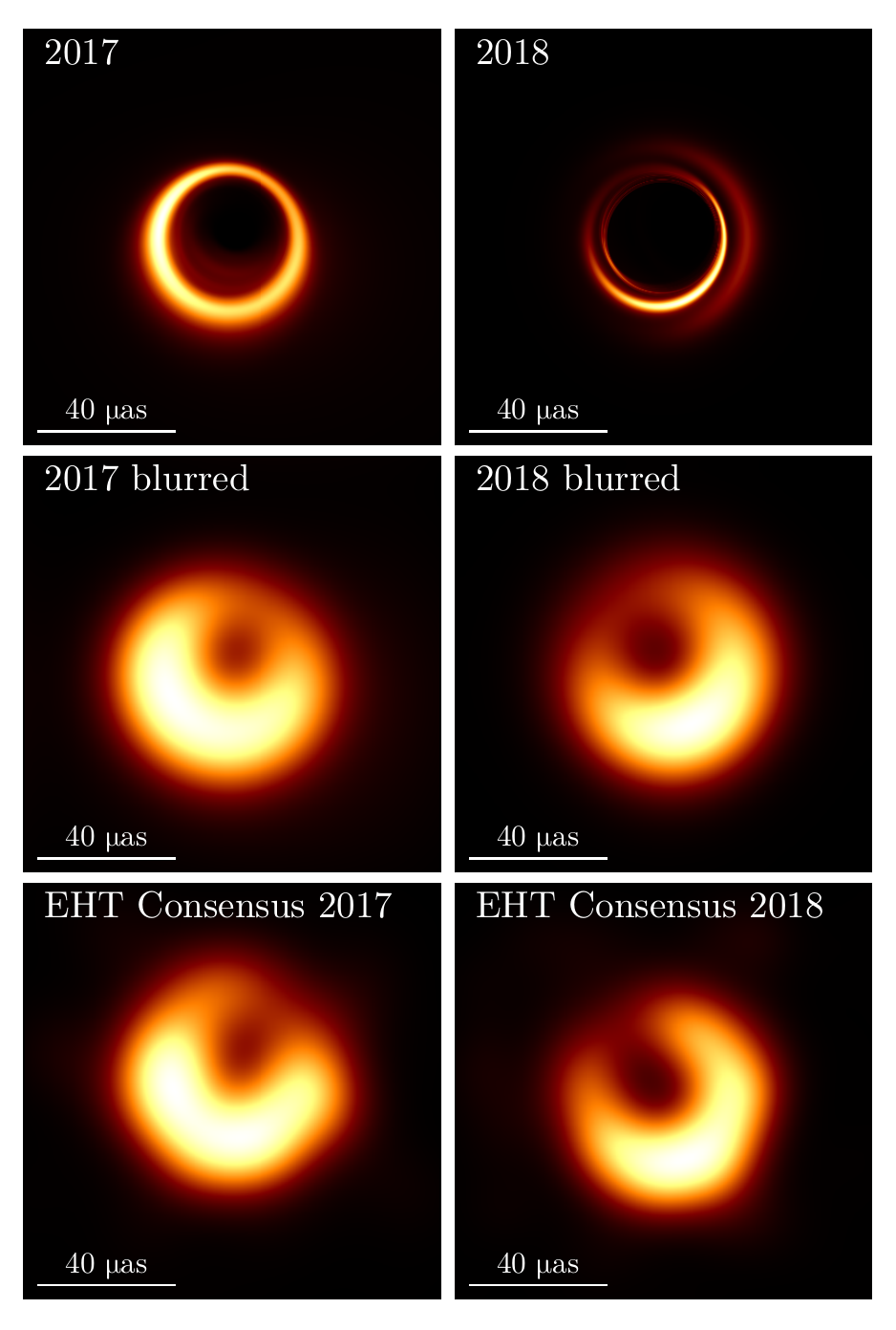}
    \caption{Images of dual-cone model from fits to the April 6, 2017 observations of M87 (left) and the April 21, 2018 observations of M87 (right).
    The top row shows the mean images of posterior chain samples at the native resolution, and the middle row shows the samples blurred to the nominal EHT resolution. For the EHT resolution, we use a Gaussian beam with a full width at half maximum (FWHM) of $20\, \mu{\rm as}$).
    The bottom row shows the corresponding consensus images from the EHT.
    }
    \label{fig:image-comparisons}
\end{figure}

\begin{figure*}[htbp]
    \centering
    \includegraphics[width=\textwidth]{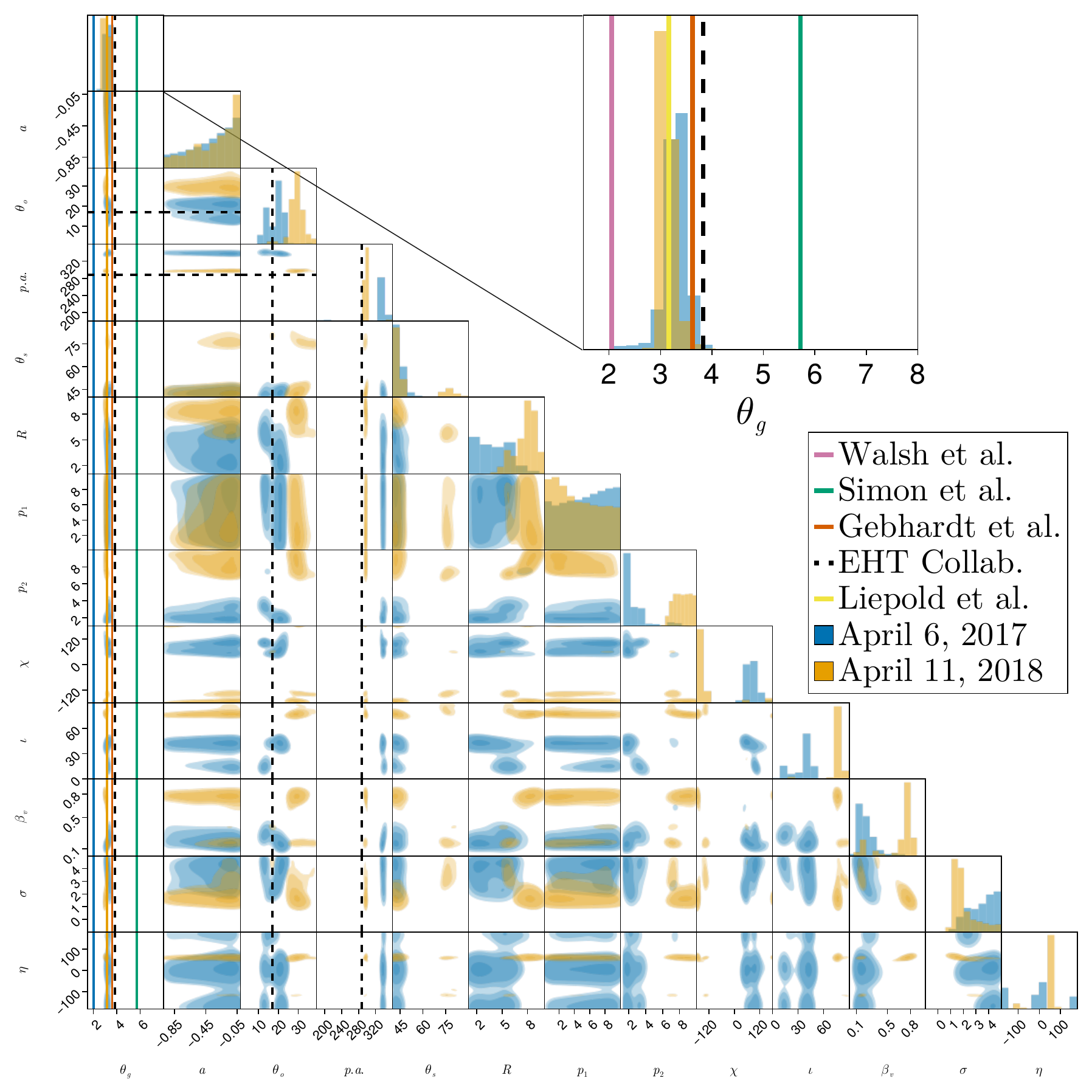}
    \caption{Full corner plot of posterior samples for the  dual-cone model fits to data from EHT observations of \m87 on April 6, 2017 (blue) and April 11, 2018 (orange).
    Vertical lines show independent mass measurements of \m87 from \citet{Gebhardt} (red), \citet{Walsh} (pink), \citet{m87VI} (black dotted), \citet{Liepold_2023} (yellow), and \citet{Simon} (green). 
    We also show the measured position angle and inclination of the large-scale jet in \m87 \citep{Walker}.
    See \autoref{tab:model_params} for a description of the model parameters.
    Both the blue and orange set of histograms have been normalized.
    }
    \label{fig:visibility-data-fit-full}
\end{figure*}

Our results are summarized in Figures \ref{fig:image-comparisons} and \ref{fig:visibility-data-fit-full}, and \autoref{tab:HDPI_EHT}.
\autoref{fig:visibility-data-fit-full} shows the full corner plot of posterior samples from the \cite{Chang_2024} analysis, which we overlay with the new posterior samples from our analysis of the 2018 data.
We also show mean images of the samples from both posteriors in \autoref{fig:image-comparisons}, which we compare to the EHTC's consensus images of the source on their respective days.
Finally, \autoref{tab:HDPI_EHT} shows the 95\% highest probability density interval (HPDI) for each of our model parameters for both the 2017 and 2018 data.

\begin{table}[ht!]
\centering
\begin{tabular}{c|c|c|c}
\hline
\hline
\multicolumn{3}{c}{95\% Highest Probability Density Intervals}\\
\hline
\textbf{Params.} & \textbf{2017} & \textbf{2018} & \textbf{EHT}\\
\hline
$\theta_{\rm g}$ & (2.84\,,\,3.75) & (2.93\,,\,3.44) & (3.05\,,\, 4.79)\footnote{From the EHTC's combined analysis of the 2017 and 2018 data \citep{M87_2018_II}.}\\
$a$ & (-0.90\,,\,-0.01) & (-0.87\,,\,-0.01) &\\
$\theta_{\rm o}$ & (11$^\circ$\,,\,24$^\circ$) & (23$^\circ$\,,\,39$^\circ$) &\\
p.a. & (200$^\circ$\,,\,347$^\circ$) & (290$^\circ$\,,\,304$^\circ$) &\\
$\theta_s$ & ($40^\circ$\,,\,$56^\circ$) & ($40^\circ$\,,\,$80^\circ$) &\\
$R$ & (1.00\,,\,8.46) & (5.23\,,\,9.72) &\\
$p_1$ & (0.71\,,\,9.99) & (0.10\,,\,9.37) &\\
$p_2$ & (1.47\,,\,7.27) & (6.77\,,\,10.00) &\\
$\chi$ & (35$^\circ$\,,\,140$^\circ$) & (177$^\circ$\,,\,239$^\circ$) &\\
$\iota$ & (10$^\circ$\,,\,49$^\circ$) & (71$^\circ$\,,\,90$^\circ$) &\\
$\beta_v$ & (0.08\,,\,0.55) & (0.15\,,\,0.83) &\\
$\sigma$ & (1.75\,,\,5.0) & (1.02\,,\,4.41) &\\
$\eta$ & (-180\,,\,174) & (-114\,,\,75) &\\
\hline
\end{tabular}
\caption{95\% highest probability density interval (HPDI) of the dual-cone model fitted to EHT observations of \m87 on April 6, 2017 and April 21, 2018.
The HPDI of the mass-to-distance ratio, black hole spin, and jet opening angle are consistent over the two year period. The large shifts between 2017 and 2018 in some parameters (e.g., $\iota$, $\chi$) indicate significant unaccounted systematic error from intrinsic variability.
}
\label{tab:HDPI_EHT}
\end{table}

We note from inspection of \autoref{fig:visibility-data-fit-full} and \autoref{tab:HDPI_EHT} that the increased data coverage resulted in stronger statistical constraints on most parameters in the 2018 fits, apart from the jet opening angle.
This behavior was expected since the 2018 data set has 339 visibilities, a $\sim$20\% increase from the 269 visibilities in the 2017 data set. 
In addition to the tighter constraints, some of the posterior distributions shift significantly between the two years.

Since the dual-cone model is expected to be representative of the average accretion flow, these shifts suggest the contribution of systematic uncertainties from some missing model component. 
Hence, the reported uncertainties are likely underestimated and some of the infered measurements could be biased.
\citet{EHTC_M87_2018} has suggested that the 2018 observations of \m87 are more representative of the average state of the source.
Indeed, we find that the dual-cone model fits the data from the 2018 observations better than those from 2017 (see \autoref{app:data_study}).

Another possibility is that some of the parameter shifts in the model are accurately tracking physical changes in the source over time. 
We can estimate how much variation we are likely to expect in some of these parameters from non-EHT observations of the source.
For instance, we expect little variation in the observed inclination and projected position angle of the spin axis since misalignments of the spin axis with the accretion flow would likely result in morphological changes in the jet on larger scales. 
The jet of \m87 is however seen to be relatively straight over multiple epochs of observations at multiple observing frequencies \citep[e.g.,][]{Walker}. 


Although we have argued that the stability of the large scale jet implies stability in the black hole spin axis,
\citet{Cui2023-tr} have observed small variations in the \m87 jet over multiple epochs, which they say could imply ${\sim}20^\circ$ shifts in the projected jet position angle over multiple years of observations. 
These shifts have been used to suggest a spin axis precession angle of $<3^\circ$ under the assumption of a rigid jet model.
If the jet is aligned with the spin axis, then the parameters inferred from fits of the dual-cone model to the 2017 observations of \m87 are inconsistent with the \citet{Cui2023-tr} study.

We thus conclude that the parameter shifts seen between the 2017 and 2018 fits are likely due to variable structures in the black hole accretion flow that are an important missing part of the model. 
Our assumption of an axisymmetric emission structure limits the variability along the azimuthal directions of the emission structures.
Relaxing this constraint would likely provide the flexibility necessary to model the missing component.

The source variability appears to have little effect on the measured mass-to-distance ratio ($\theta_{\rm g}$), which has comparable constraints in both years despite striking changes in the images of the dual-cone model (\autoref{fig:image-comparisons}). This consistency indicates that the inferred values for $\theta_{\rm g}$ may be robust under source variation. The spin ($a$) and the inner shape of the emission geometry closer to the horizon ($p_1$) are unconstrained in both years. 
Our inability to constrain these parameters suggests that measurements of the black hole spin and inner shadow \citep{Chael_2021} will be difficult at the current EHT resolution and dynamic range.

In contrast, we see notable shifts in the emissivity profile ($R$ and $p_2$), the fluid speed ($\beta_v$), the fluid direction ($\chi$), the magnetic field orientation ($\iota$ and $\eta$), the spin-axis inclination ($\theta_o$), and the spin axis position angle ($p.a.$).
These shifts imply a strong dependence of the inferred accretion state of the model on source variability.  

Changes in the inferred fluid flow direction and projected spin axis position angle are particularly emblematic of the effects of variability.
Images and model fits to the EHT observations of \m87 that are external to this study all feature a peak brightness position angle shift from 2017 to 2018 \citep[][]{M87_2018_I}.
Our model also identifies a shift in the peak brightness position angle, which is produced from a shift in the inferred accretion flow direction from retrograde in 2017 (since the sign of $\chi$ is opposite to the sign of $a$) to a prograde flow in 2018 data.
The change from a retrograde to a prograde flow is unlikely given the stability of the jet; hence, this parameter variation is more likely to be a consequence of model misspecification.

\autoref{fig:image-comparisons} illustrates how model misspecification is absorbed into the dual-cone model parameters. 
The top row of the figure shows the mean images of posterior samples from the 2017 and 2018 data analyses, while the middle row shows these images blurred to the nominal EHT resolution. 
The bottom row shows the EHT consensus images for comparison. 
Since the dual-cone model is a representation of the average accretion flow, any variation between images in 2017 and 2018 is likely to be caused by the model parameters incorrectly attributing variable structure to the time-averaged image. 
The model successfully reproduces the striking changes between the image morphology in these two epochs, particularly in the shift of peak brightness around the ring. 
This consistency suggests that the model absorbs variability into parameters of the time-averaged model. 
We also note the presence of persistent substructure in the 2018 mean images. 
This substructure is much smaller than the EHT's nominal beam size, and its consistency across posterior samples suggests that our model underfits the image structure of the source (see the top row of \autoref{fig:image-comparisons}).
Thus, we conclude that many parameters from fitting the dual-cone model to individual data snapshots may not be representative of the source average.


We also note the shift in the HPDI of the observer inclination and the projected angle of the spin axis on the observer's screen in 2018 when compared to 2017 (see \autoref{tab:HDPI_EHT}).
The HPDI of the measured $p.a.$ from the 2018 data has changed to no longer be in tension with the measured $p.a.$ of the $7$mm jet.
Similarly, the HPDI of $\theta_o$ from the 2018 data is also different from its 2017 range, where it now stands in weak tension with the measured $17^\circ$ inclination of the large-scale jet seen at 7\,mm wavelength \cite{Walker}.
In general, the dual-cone model provides a better fit to the 2018 data than the 2017 data.

\section{Summary}
\label{sec:summary}

We have fit a semi-analytic model to EHT observations of \m87 in 2017 and 2018. This model, introduced by \citet{Chang_2024}, has 13 parameters that characterize the black hole, observer location, and emission region. The low dimensionality and differentiability of this model allow it to be used in a Bayesian inference framework that efficiently samples the full posterior distribution for EHT observations. This approach provides a powerful complement to the EHTC analyses, allowing us to estimate posterior distributions for the black hole mass-to-distance ratio and spin that marginalize over the unknown emission geometry.  

Our model successfully reproduces the striking differences in the EHT images in 2017 and 2018. However, because the model is designed to describe the time-averaged emission structure, this success also indicates that the model parameters are systematically biased by the (unmodeled) intrinsic variability. Fits of our dual-cone model to single epochs of EHT data in 2017 and 2018 result in consistent mass-to-distance ratio and spin inferences. However, discrepancies in other model parameters indicate the presence of systematic uncertainties from intrinsic source variability. 

Our approach provides a pathway to parameter inference from EHT data, using multiple epochs to quantify systematic uncertainty from intrinsic variability. Future extensions could integrate variable emission structure directly into the underlying model. While we have focused on the two independent realizations of \m87 that can be studied using the EHT, another interesting application of this method is for EHT studies of \sgra. With a gravitational timescale of only $GM/c^3 \approx 20\,{\rm seconds}$, observations of \sgra sample many independent statistical realizations within a single night, providing sharper estimates of the systematic uncertainty from intrinsic variability. These approaches will be vital for connecting observations with expansions of the EHT on the ground \citep{Doeleman_2023,EHT_midrange} or with a space-enhanced EHT through the Black Hole Explorer \citep[BHEX;][]{BHEX_Concept,BHEX_Instrument}, to achieve precise and accurate measurements of the black hole parameters for both \m87 and \sgra.


\begin{acknowledgments}
\section{Acknowledgements}
We thank Rohan Dahale, Boris Georgiev, Britton Jeter and Hung-Yi Pu for their insightful discussions. We acknowledge financial support from the National Science Foundation (AST-2307887). This publication is funded in part by the Gordon and Betty Moore Foundation, Grant GBMF12987. This work was supported by the Black Hole Initiative, which is funded by grants from the John Templeton Foundation (Grant \#62286) and the Gordon and Betty Moore Foundation (Grant GBMF-8273) - although the opinions expressed in this work are those of the author(s) and do not necessarily reflect the views of these Foundations.
\end{acknowledgments}

\pagebreak
\appendix

\section{Quality of Data Fits}
\label{app:data_study}

\begin{figure}[ht]
    \centering
    \includegraphics[width=\linewidth]{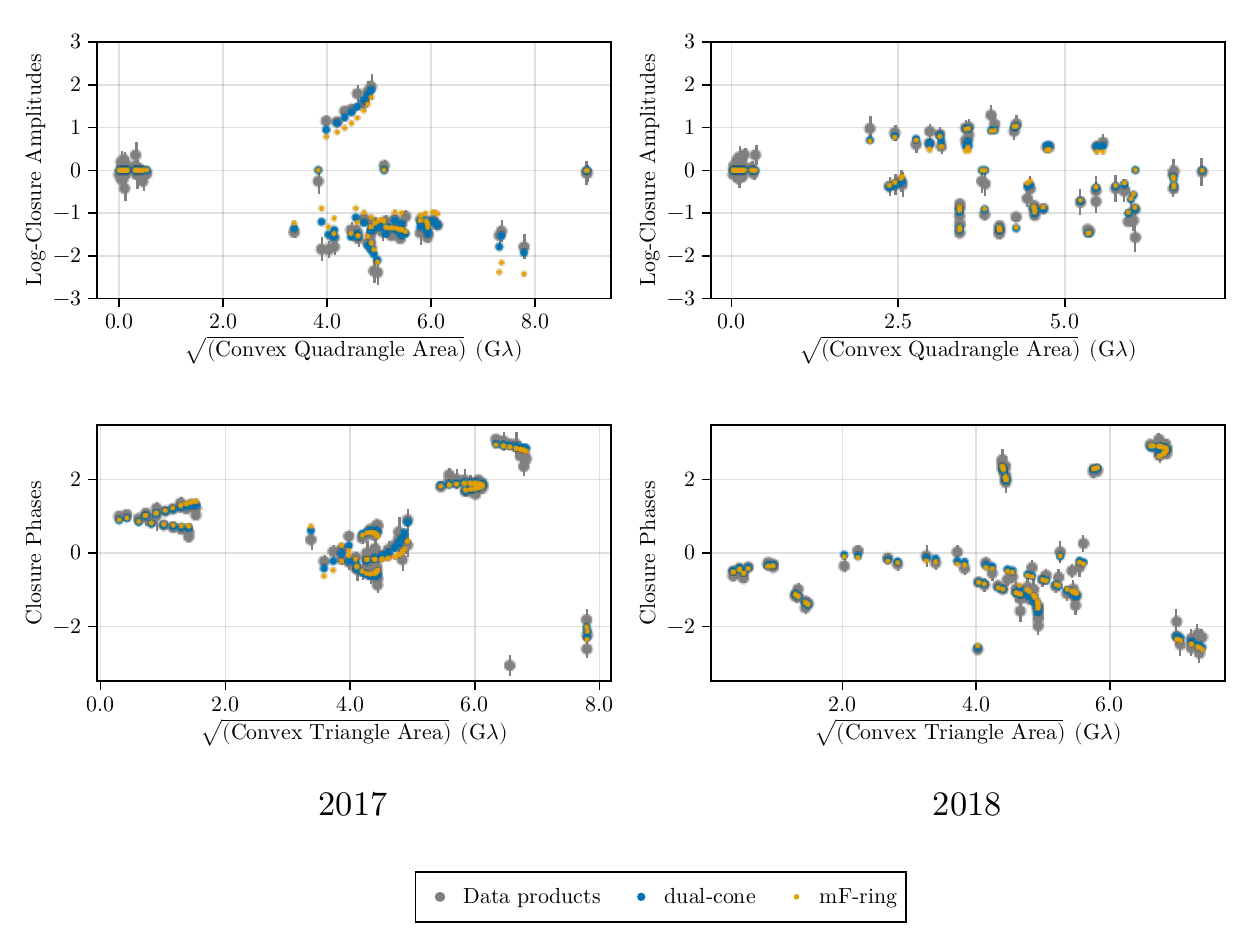}
    \caption{Best fits (i.e., the MAP estimates) for the dual-cone model (blue) and the fiducial mF-ring model (orange) to log-closure amplitudes (top row) and closure phases (bottom row) constructed from data acquired during observations of \m87 by the EHT (grey).
    The fits to the the 2017 and 2018 data products are shown in the left and right columns, respectively.
    }
    \label{fig:map-comparisons}
\end{figure}
\begin{figure}
    \centering
    \includegraphics[width=0.75\linewidth]{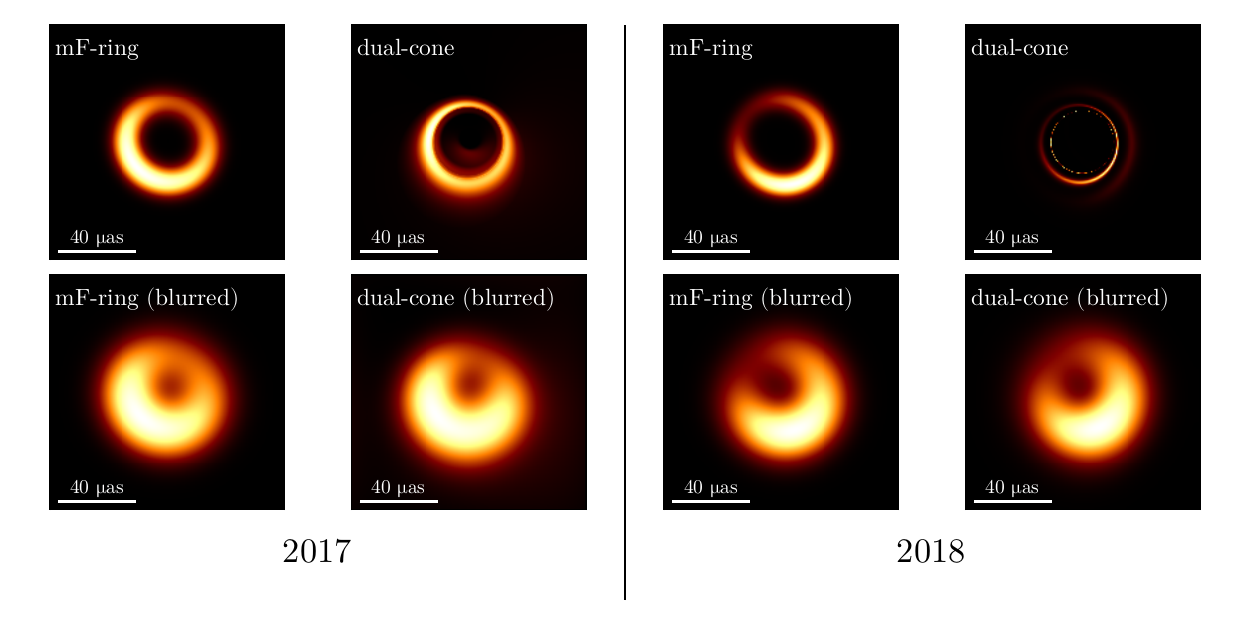}
    \caption{Images of the MAP estimates of the mF-ring model and the dual-cone model for fits to EHT observations of \m87 in 2017 (left; center-left) and in 2018 (center-right; right).
    The top row shows images at their native resolutions; the bottom row shows images blurred to the nominal EHT resolution. For the latter, we use a Gaussian beam of 20 $\mu as$ FWHM.
    }
    \label{fig:map-images}
\end{figure}
\begin{figure}
    \centering
    \includegraphics[width=0.75\linewidth]{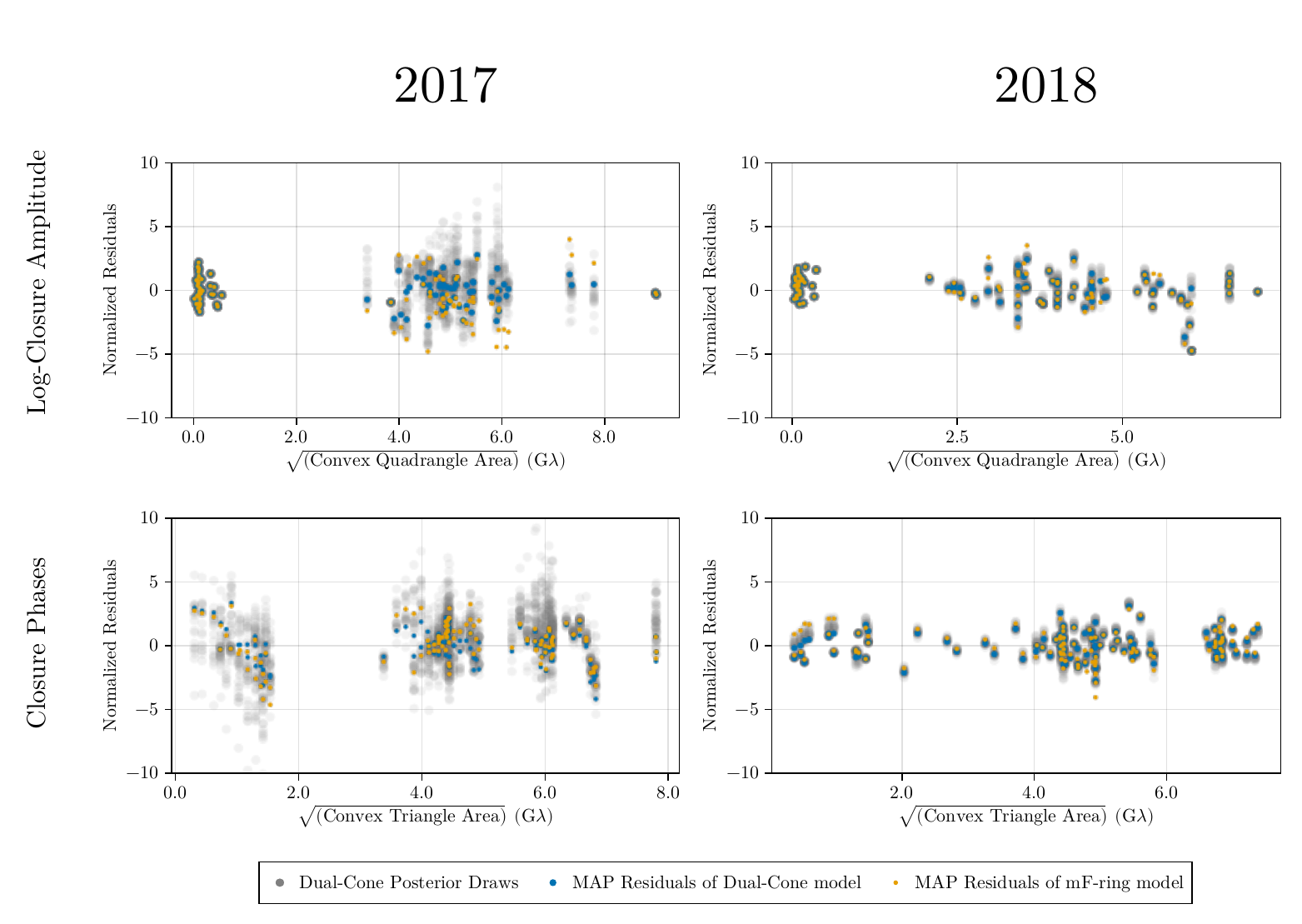}
    \caption{Normalized log-closure amplitude (left column) and closure phase residuals (right column) of mF-ring and dual-cone model fits to the 2018 observations of \m87 by the EHT.
    Each panel shows residuals of posterior samples (grey) and the maximum a-posteriori estimate (blue).
    Residuals of posterior samples from fits of the dual-cone model to 2017 and 2018 observations of \m87 are shown in the top row and center row respectively.
    Similarly, residuals of posterior samples from fits of the mF-ring model to the 2018 observations of \m87 are shown in the bottom row.
    }
    \label{fig:residual-plots}
\end{figure}

This section summarizes quality-of-fit studies of the dual-cone model to the data products used in this study. 
We use the fiducial geometric model, which we will call the mF-ring model hereafter, from the \citet{M87_2018_I} analysis of the 2018 \m87 data as a comparator to assess the quality-of-fit of the dual-cone model.
We show the maximum a-posterior (MAP) fits of both models to the 2017 and 2018 data products in \autoref{fig:map-comparisons}, and their images in \autoref{fig:map-images}.
We show the normalized residuals of posterior samples taken from the dual-cone model fits in \autoref{fig:residual-plots}, where we overlay the residuals from the MAP estimates from the dual-cone model and mF-ring model fits for comparison.


The mF-ring model is a 9-parameter model that was chosen as a fiducial geometric model of \m87 in the \citet{M87_2018_I} 2018 analysis.
This model was chosen from a set of candidates used by the collaboration to describe the image morphology of the source, and was the best performing model based on a Bayesian evidence criterion.
We note that the mF-ring model and the dual-cone model result in similar fit qualities to the data in 2017 and 2018 (see \autoref{fig:map-comparisons} and \autoref{fig:residual-plots}).
This similarity suggests that visual differences between the two models are likely indistinguishable at the current resolution provided by the EHT array.
The similar structure seen in images of the models when blurred to the nominal EHT resolution in \autoref{fig:image-comparisons} also suggests that the data is not sufficiently constraining to distinguish between the two.

Both the dual-cone model and the mF-ring model provide better overall fits to the source in 2018 than in 2017 as evidenced by the differences in the spread in residuals between the two years.
This behavior is consistent with our interpretation that the dual-cone model is representative of the average state of the system, since studies of GRMHD simulations of the source suggest that the 2018 observations of \m87 provide a snapshot that is more representative of its average accretion flow \citep[][]{M87_2018_I, M87_2018_II}.
\bibliography{references}
\end{document}